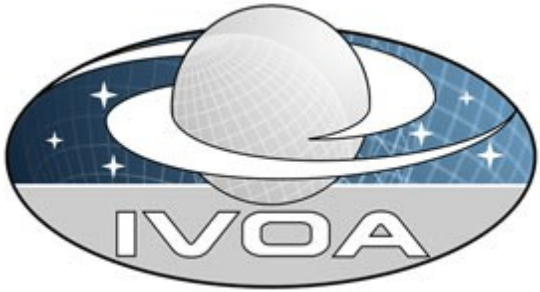

*I*nternational

*V*irtual

*O*bservatory

*A*lliance

# Table Access Protocol
# Version 1.0
## IVOA Recommendation 2010 March 27

**Interest/Working Group:**

http://www.ivoa.net/cgi-bin/twiki/bin/view/IVOA/IvoaDAL

**This version:**

REC-TAP-1.0

**Latest version:**

http://www.ivoa.net/Documents/TAP/

**Previous version(s):**

http://www.ivoa.net/Documents/TAP/20100225/

http://www.ivoa.net/Documents/TAP/20091225/

http://www.ivoa.net/Documents/TAP/20091006/

http://www.ivoa.net/Documents/TAP/20090608/

**Authors:**

P. Dowler, G. Rixon, D. Tody

**Editors:**

P. Dowler

**Contributors:**

K. Andrews, J. Good, R. Hanisch, G. Lemson, T. McGlynn, K. Noddle, F. Ochsenbein, I. Ortiz, P. Osuna, R. Plante, J. Salgado, A. Stebe, A. Szalay

## Abstract

The table access protocol (TAP) defines a service protocol for accessing general table data, including astronomical catalogs as well as general database tables. Access is provided for both database and table metadata as well as for actual table data.   This version of the protocol includes support for multiple query languages, including queries specified using the Astronomical Data Query Language (ADQL [1]) and the Parameterised Query Language (PQL, under development) within an integrated interface. It also includes support for both synchronous and asynchronous queries.  Special support is provided for spatially indexed queries using the spatial extensions in ADQL. A multi-position query capability permits queries against an arbitrarily large list of astronomical targets, providing a simple spatial cross-matching capability.   More sophisticated distributed cross-matching capabilities are possible by orchestrating a distributed query across multiple TAP services.

## Status of This Document

This document has been produced by the Data Access Layer Working Group.

It has been reviewed by IVOA Members and other interested parties, and has been endorsed by the IVOA Executive Committee as an IVOA Recommendation. It is a stable document and may be used as reference material or cited as a normative reference from another document. IVOA's role in making the Recommendation is to draw attention to the specification and to promote its widespread deployment. This enhances the functionality and interoperability inside the Astronomical Community.

A list of current IVOA Recommendations and other technical documents can be found at http://www.ivoa.net/Documents/.

## Contents

# 1 Introduction (informative)

The *Table Access Protocol* (TAP) is a Web-service protocol that gives access to collections of tabular data referred to collectively as a *tableset*. TAP services accept queries posed against the *tableset* available via the service and return the query response as another table, in accord with the relational model. Queries may be submitted using various query languages and may execute synchronously or asynchronously. Support for the Astronomical Data Query Language (ADQL, [1]) is mandatory; support for other query languages such as Parameterised Query Language (PQL, under development) or native SQL is optional.

The result of a TAP query is another table, normally returned as a VOTable. Support for VOTable output is mandatory; all other formats are optional.

The table collections made accessible via TAP are typically stored in relational database management systems (RDBMS). A TAP service exposes the database schema to client applications so that queries can be posed directly against arbitrary data tables available via the service.

Multi-table operations such as joins or cross matches are possible provided the tables are all managed by the local TAP service, and provided the service supports these capabilities. Larger scale operations such as a distributed cross match are also possible, but require combining the results of multiple TAP services.

## 1.1 Query Types

TAP services support three kinds of queries: data queries, metadata queries, and Virtual Observatory Support Interface (VOSI [6]) queries.

### 1.1.1 Data Queries

Data queries apply to the astronomical content served by a TAP service. This is the reason for providing a TAP service. All the other kinds of query support the ability to make data queries. Data queries may be specified in any query language supported by the service.

### 1.1.2 Metadata Queries

Metadata queries work like data queries, using the same query languages, but they are applied to standardized tables that are a subset of, and patterned after, *information schema* in RDBMS; the content of these tables explains the data model of a particular TAP service. Metadata queries allow a client to discover the names of tables and columns to be used in data queries.

### 1.1.3 VOSI

The Virtual Observatory Support Interface (VOSI [6]) specifies the base service interface common to all VO services. VOSI requests supply metadata concerning the availability ('VOSI-availability') of a TAP service, its main interfaces ('VOSI-capabilities'), and its data model ('VOSI-tables'). VOSI-capabilities and VOSI-tables outputs use the same XML schema [7] as the IVOA registry [16] and can be incorporated in service registrations.

## 1.2 Query Languages

TAP supports the use of multiple query languages, some of which are described here.

### 1.2.1 ADQL Queries

Support for ADQL queries is mandatory. ADQL can be used to specify queries that access one or more tables provided by the TAP service, including the standard metadata tables. In general, the client must access table metadata in order to discover the names of tables and columns and then formulate queries. ADQL queries provide a direct (low-level) access to the tables; a query will be written for a specific TAP service and will not be usable with other services unless the query refers only to common tables and columns. It is also possible that the service registration (in an IVOA Registry) may include sufficient table metadata to enable queries to be written directly.

Details of the ADQL language may be found in [1].

### 1.2.2 PQL Queries

Support for PQL is optional. PQL can be used to formulate queries that access a single table provided by the TAP service, including the standard set of metadata tables. PQL can also be used in some cases without first querying the metadata tables by using the PQL parameters which carry sufficient meaning to enable the service to decide which tables and columns to use (e.g. POS, SIZE, REGION, BAND, TIME).

Details of the PQL language (parameters) are not part of the TAP specification.

### 1.2.3 Other Query Languages

A TAP service may also support use of other query languages, including pass-through of native SQL directly to an underlying DBMS, by describing such capabilities in the service metadata and allowing custom values of the service parameters. This mechanism allows future developments within the VOQL Working Group and outside the IVOA to be used without revising the TAP specification.

## 1.3 Query Execution

The TAP service specification defines both synchronous and asynchronous query execution. Users select synchronous or asynchronous execution by chosing the appropriate resource below the base URL for the service (see 2.2 ). A query is synchronous if the results of the query are delivered in the HTTP response to the request that originally posed the query. If the service returns an immediate HTTP-response upon accepting a query and the client later obtains the results of the query in response to a separate HTTP request, then we say the request is asynchronous.

### 1.3.1 Asynchronous Queries

Asynchronous query support is mandatory. Asynchronous queries require that client and server share knowledge of the state of the query during its execution and between HTTP exchanges. They are an example of stateful interactions. In TAP, the mechanism by which the clients and services share the state of transactions is based on the Universal Worker Service (UWS) pattern [3].

### 1.3.2 Synchronous Queries

Synchronous query support is mandatory. Synchronous queries execute immediately and the client must wait for the query to finish. If the HTTP request times out or the client otherwise loses the connection to the service before receiving the response, then the query fails.

Synchronous query execution is adequate when the query will execute quickly and with a small number of results, or when they can at least start returning results quickly. They are generally simple to implement using standard web technologies and easy to use from a browser or scripting environment. However, synchronous requests are generally not sufficient and are likely to fail for queries that take a long time to execute, especially before returning any results.

## *1.4 Interface Overview*

Table Access Protocol (TAP) is implemented over the HTTP protocol using standard HTTP GET and POST requests and conventions. A TAP request specifies one or more parameter key/value pairs; both keys and values are strings. The keys used are discussed in this specification and in the specifications for query languages supported by a service. The values may need to be encoded, using standard URL-encoding to replace non-alphanumeric characters with %xx sequences. For the following examples, http://example.com/tap/ is the base URL for a TAP service.

This is an example[1] of a synchronous ADQL query on *r* magnitude:

```
HTTP POST htp://example.com/tap/sync
REQUEST=doQuery
LANG=ADQL
QUERY=SELECT * FROM magnitudes as m where m.r>=10 and m.r<=16
```

The equivalent PQL query would be:

```
HTTP POST htp://example.com/tap/sync
REQUEST=doQuery
LANG=PQL
```

---

1  Throughout this document, example URL parameters are shown without URL-encoding for clarity; ADQL queries in particular must be encoded since they have spaces and various special characters.

```
FROM=magnitudes
WHERE=r,10/16
```

Synchronous queries return the table of results in the HTTP response[2] to the initial request. In the examples above, the output format defaults to VOTable; the FORMAT parameter could be added to select a different format.

Asynchronous queries are started in the same way as the synchronous kind, using the /async endpoint:

```
HTTP POST http://example.com/tap/async
REQUEST=doQuery
LANG=ADQL
QUERY=SELECT * FROM magnitudes AS m WHERE m.r>=10 AND m.r<=16
```

or

```
HTTP POST http://example.com/tap/async
REQUEST=doQuery
LANG=PQL
FROM=magnitudes
WHERE=r,10/16
```

The service's response to these requests is a URL representing the query's state and progress and where the state may be monitored and controlled. The query result or an error document can then be retrieved from a URL associated with the job. This is an application of the UWS pattern.

Positional queries have special support in PQL. This is a cone search on a specified table:

```
HTTP POST http://example.com/tap/sync
REQUEST=doQuery
LANG=PQL
POS=12,34
SIZE=0.5
FROM=foo
```

In addition to the sync and async resources for query execution, a TAP service also has metadata resources defined by the VOSI standard. The availability of a service can be monitored by accessing:

```
HTTP GET http://example.com/tap/availability
```

See 2.2.3 for details of the availability resource.

The complete table metadata can be obtained:

```
HTTP GET http://example.com/tap/tables
```

See 2.2.5 for details of the table metadata resource.

The capabilities can be obtained by:

```
HTTP GET http://example.com/tap/capabilities
```

---

2   Synchronous requests may issue a redirect to the result using HTTP code 303: See Other.

The capabilities are also accessible via a service request to the synchronous query resource:

```
HTTP GET http://example.com/tap/sync
REQUEST=getCapabilities
```

This output lists support for optional TAP functionality and additional implemented interfaces. See  2.2.4  for details.

# 2 Requirements for a TAP service (normative)

The keywords "must", "required", "should", and "may" as used in this document are to be interpreted as described in the W3C specifications (IETF RFC 2119 [2]). Mandatory interface elements are indicated as **must**, recommended interface elements as **should**, and optional interface elements as **may** or simply "may" without the bold face font.

The TAP standard is identified using standardID="ivo://ivoa.net/std/TAP". This document specified version 1.0 of the standard.

## 2.1 Feature Overview

An implementation of a TAP service provides features as follows.

| Feature | Resource | Resource support | Parameters | Parameter support |
|---|---|---|---|---|
| synchronous query execution | /sync | **must** | | |
| asynchronous query execution | /async | **must** | | |
| availability (VOSI) | /availability | should | | |
| capabilities (VOSI) | /capabilities | **must** | REQUEST=getCapabilities | **must** |
| table metadata (VOSI) | /tables | should | | |
| ADQL queries | | | REQUEST=doQuery LANG=ADQL | **must** |
| PQL queries | | | REQUEST=doQuery LANG=PQL | may |
| other query languages | | | REQUEST=doQuery LANG=<other> | may |
| table upload | | | UPLOAD | may |
| VOTable output | | | FORMAT | **must** |
| other formats | | | FORMAT | should |
| limiting output | | | MAXREC | **must** |
| logging | | | RUNID | should |

The resources and parameters are described in detail below. The description of these resources and parameters spell out how the requirements here are to be implemented.

TAP service registration in the IVOA resource-registry is specified in section 3 .

## 2.2 Web resources

A TAP service **must** be represented as a tree structure of web resources each addressable via a URL in the http scheme, or the https scheme, or both.

The web resource at the root of the tree **must** represent the service as a whole. This specification defines no standard representation for this root resource. Implementations **may** provide a representation, or may return a '404 not found' response to requests for the root web-resource. One possible representation is an HTML page describing the scientific usage and content of the service. TAP clients **must not** depend on a specific representation of the root web-resource.

### 2.2.1 /sync

A TAP service **must** provide a web resource with relative URL */sync* that is a direct child of the root web resource. This web resource represents the results of synchronous requests. The exact form of the query, and hence the representation of the resource, is defined by the query parameters as listed in section 2.3. Representations of results of queries and VOSI outputs are defined in sections 2.7.1 and 2.7.2 respectively.

For query languages that produce a single result (e.g. ADQL) executed using the /sync endpoint, the result of a successful query is returned in the response or the response includes an HTTP redirect (303: See Other) to a resource (uri) from which the result may be retrieved.

An HTTP-GET request to the */sync* web resource **may** return a cached copy of the representation. This cached copy might come from an HTTP cache between the client and the service, and the service **may** also maintain its own cache. Clients which require an up-to-date representation of volatile data or metadata must use HTTP POST.

### 2.2.2 /async

A TAP service **must** provide a web resource with relative URL */async* that is a direct child of the root web resource. This web resource represents controls for asynchronous queries. Specifically, the web resource **must** represent the job-list as specified in the UWS standard [3].

The child web resources of the /async resource are as specified by UWS. These are descendants of the */async* web-resource, and they include a web resource that represents the eventual result of an asynchronous query, e.g.:

```
http://example.com/tap/async/42/results/result
```

where the base URL for the TAP service is:

```
http://example.com/tap
```

the UWS job list is:

```
http://example.com/tap/async
```

and the job resource is

```
http://example.com/tap/async/42
```

where 42 is the job identifier. A client making an asynchronous request must use the UWS facilities to monitor or control the job. In addition to the job list and job resource above, UWS specifies the name and semantics of the a small set of child resources used to view and control the job, e.g.:

```
http://example.com/tap/async/42/phase
http://example.com/tap/async/42/quote
http://example.com/tap/async/42/executionduration
http://example.com/tap/async/42/destruction
http://example.com/tap/async/42/error
http://example.com/tap/async/42/parameters
http://example.com/tap/async/42/results
http://example.com/tap/async/42/owner
```

Successful TAP queries produce results which **must** be accessible as resources under the UWS result list, e.g.:

```
http://example.com/tap/async/42/results/
```

Failed TAP queries produce an error document (see 2.9 ) which **must** be accessible as the error resource, e.g.:

```
http://example.com/tap/async/42/error
```

For query languages that produce a single result executed using the /async endpoint, the result of a successful query can be found within the result list specified by UWS [3]; the result must be named *result* and thus clients are able to access it directly, e.g.:

```
http://example.com/tap/async/42/results/result
```

Access of this resource must deliver the result, either directly or as an HTTP redirect (303: See Other) to a resource from which the result may be retrieved.

For query languages that may produce multiple result resources, the names of the results are not specified (they may be specified in the specification for the language). The client can always access the result list resource as specified by UWS [3].

If the query returned no rows, the result resource **must** exist and contain no data rows. Details on interacting with these resources are specified in the UWS standard; for examples specific to TAP see Section 5 below.

### 2.2.3 /availability

The VOSI availability metadata **should** be accessible from a web resource with relative URL /availability that is a direct child of the root web resource. If implemented, the /availability resource **must** be accessible via the http GET method. The content is described by [6].

Services which do not implement the /availability resource **must** respond with an HTTP response code of 404 when this resource is accessed.

### 2.2.4 /capabilities

The service capabilities **must** be accessible from a web resource with relative URL /capabilities that is a direct child of the root web resource. The /capabilities resource must be accessible via the http GET method. The content is described by [8].

### 2.2.5 /tables

The table metadata **should** be accessible from a web resource with relative URL /tables that is a direct child of the root web resource. The /tables resource must be accessible via the http GET method.  The content is described by [7] and is equivalent to the metadata from the TAP_SCHEMA described in  2.6 .

Services which do not implement the /tables resource **must** respond with an HTTP response code of 404 when this resource is accessed.

## 2.3 Parameters for HTTP requests

The */sync* and */async* web-resources **must** accept the parameters listed in the following sub-sections. In a synchronous request, the parameters select the representation returned in the response message. In an asynchronous request, the parameters select the representation of the eventual query result rather than the response to the initial request.

Requirements on the presence and values of parameters described below are enforced only when the TAP request is executed (not when individual HTTP requests are handled). Thus, for asynchronous TAP queries, the parameter requirements must be satisfied (and errors returned if not) only when the query is run in (in the sense of UWS job execution). Specifically, asynchronous queries may be created with with no parameters and multiple, subsequent HTTP POST actions may specify the parameters in any order.

Not all combinations of the parameters are meaningful. For example, if a request carries  LANG=*ADQL* then the *SELECT* parameter (from PQL) is spurious. If a service receives a spurious parameter in an otherwise correct request, then the service **must** ignore the spurious parameter, must respond to the request normally and **must not** report errors concerning the  spurious parameter.

### 2.3.1 REQUEST

This parameter distinguishes current service operations, makes it possible to extend the service specification (with additional or custom operations), and specifies how other parameters should be interpreted. If a TAP service attempts to execute a TAP request without this parameter or with an incorrect value for this parameter, then the service **must** reject the request and return an error document as the result.

These are the standard values of the parameter:

- doQuery: execute a query
- getCapabilities: return VOSI-capabilities metadata

All requests to execute (/async or /sync) a query using a query language **must** include REQUEST=doQuery and **must** include the LANG parameter. For other values of REQUEST, additional parameters may or may not be required. The REQUEST=getCapabilities service operation **must** be supported for synchronous (/sync) requests and is not defined for asynchronous (/async) requests.

For synchronous queries, the HTTP request **must** also include additional parameters (see below) with the details of the query. These are used for metadata queries and data queries.

For asynchronous queries, the additional parameters may be included with the HTTP request that creates the query (the UWS job) or they may be POSTed directly to the created job resource, in one or more separate HTTP requests. The parameter names remain the same in both cases.

### 2.3.2 VERSION

The *VERSION* parameter specifies the TAP protocol version number. The format of the version number, and version negotiation, are described in section 2.9.2 .

A TAP service **must** support the *VERSION* parameter. This specification is for TAP 1.0; the client would specify VERSION=1.0 if the request is made using the protocol described here.

### 2.3.3 LANG

The LANG parameter specifies the query language. The service **must** support *LANG* and the client **must** provide a value with REQUEST=doQuery. The only standard values for the *LANG* parameter are ADQL (a required language) and PQL (reserved value for an optional language which is under development). Support for other languages and the *LANG* value to use with them is described in the service capabilities.

For example, an ADQL query would be performed with

```
REQUEST=doQuery
LANG=ADQL
QUERY=<ADQL query string>
```

A PQL query would be performed with

```
REQUEST=doQuery
LANG=PQL
<PQL-specific parameters>
```

The value of LANG is a string specifying the language and optionally the language version used for the query parameter(s), as defined by the service capabilities. The client **may** specify the version of the query language, e.g. LANG=ADQL-2.0 (the syntax should be as shown) or it may omit the version, e.g. LANG=ADQL. The service **should** return an "unknown query language" error as described in 2.9 if an unsupported language or an incompatible language version is specified.

### 2.3.4 QUERY

The QUERY parameter is used to specify the ADQL query. It may also be used to specify the query for other values of LANG (e.g. LANG=<some RDBMS-specific SQL variant>) which are not specified in this document but may be described in the service capabilities.

A service **must** support the *QUERY* parameter because ADQL is a required language. The case sensitivity of the query string is defined solely by the query language specification. In the case of ADQL 2.0, for example, the query is not case sensitive except for character literals; schema, table, and column names, function names, and other ADQL keywords are not case sensitive.

Within the ADQL query, the service **must** support the use of timestamp values in ISO8601 format, specifically yyyy-MM-dd['T'HH:mm:ss[.SSS]], where square brackets denote optional parts and the 'T' denotes a single character separator (T) between the date and time parts.

If the tables that are queried through a service contain columns with spatial coordinates and the service supports spatial querying via the ADQL "region" constructs, the service **must** support the *INTERSECTS* function and it **must** support the following geometry functions: *REGION*, *POINT*, *BOX*, *CIRCLE*, *COORD1*, *COORD2*, *COORDSYS*. Support for the *AREA*, *CONTAINS*, and *POLYGON* functions are optional. If the service supports the *REGION* function, it **must** support region encoding in STC-S format (see section 6 ); the extent of STC-S support within the *REGION* function is left up to the implementation. Coordinate system specification for *POINT*, *BOX*, *CIRCLE*, and *POLYGON* **must** use values from STC-S as described in section 6 .

> *Note: Although it is allowed by the ADQL syntax, clients should be careful when mixing constants and column references for coordinate system and coordinate values. For example, POINT('ICRS', t.ra, t.dec) does not cause t.ra and t.dec to be transformed to ICRS; it simply tells the service to treat the values as being expressed in that coordinate system.*

### 2.3.5 Parameters for PQL

A number of parameters will be defined by the PQL standard for use in parametric queries. All of the parameters for PQL are are used unchanged in TAP.

Within the PQL query, the service **must** support the use of timestamp values in ISO8601 format (see 2.3.4 ).

If the table that is queried contains columns with spatial coordinates and the service provider wants to enable the caller to perform spatial queries, the service **must** support the PQL spatial constraint parameters (POS,SIZE and REGION). If a service supports the REGION parameter, it **must** support region encoding in STC-S as decribed in section  6 ; the extent of STC-S support within the *REGION* function is left up to the implementation. Coordinate system qualifiers **must** use values from from STC-S as described in section  6 .

PQL defines symbolic values (@something). In TAP these can be used to refer to  an uploaded table (see  2.5 ) with parameters that support a table reference. When used in this way, the uploaded table must be treated as if in the TAP_UPLOAD schema (e.g. @TAP_UPLOAD.mytable). Details on how to use table references in PQL will be described in the PQL specification.

### 2.3.6 FORMAT

The *FORMAT* parameter indicates the client's desired format for the table of results of a query. Its value **should** be a MIME type for tabular data or one of the following shorthand forms:

| table type | MIME type(s) | short form |
|---|---|---|
| VOTable | application/x-votable+xml text/xml | votable |
| comma separated values | text/csv | csv |
| tab separated values | text/tab-separated-values | tsv |
| FITS binary table | application/fits | fits |
| pretty-printed text | text/plain | text |
| pretty-printed Web page | text/html | html |

Both MIME types and the shorthand forms are insensitive to case. If the *FORMAT* parameter is omitted, the default format is VOTable.

A TAP service **must** support VOTable as an output format, **should** support CSV and TSV output and **may** support other formats. A TAP service **must** accept a *FORMAT* parameter indicating a format that the service supports and **should** reject queries where the *FORMAT* parameter specifies a format not supported by the service implementation.

### 2.3.7 MAXREC

The service **must** accept a *MAXREC* parameter specifying the maximum number of table records (rows) to be returned. If *MAXREC* is not specified in a query, the service **may** apply a default value or **may** set no limit. If the result set for a query exceeds this value, the service **must** only return the requested number of rows. If the result set is truncated in this fashion, it must include an overflow indicator as specified in section  2.7.4 .

The service **must** support the special value of *MAXREC=0.* This value indicates that, in the event of an otherwise valid request, a valid output table be returned containing metadata, no table data rows, and an overflow indicator as specified in section 2.7.4 . The service is not required to execute the query; a successful MAXREC=0 request does not necessarily mean that the query is valid and the overflow indicator does not necessarily mean that there is at least one row satisfying the query. The service **may** perform validation and may try to execute the query, in which case a MAXREC=0 request can fail.

A query with *MAXREC=0* can be used with a simple query (e.g. SELECT * FROM some_table) to extract and examine the VOTable metadata (assuming *FORMAT=votable*). Note: in this version of TAP, this is the only mechanism to learn some of the detailed metadata, such as coordinate systems used.

### 2.3.8 RUNID

The service **should** implement the *RUNID* parameter, used to tag service requests with the job ID of a larger job of which the request may be part. The *RUNID* parameter is defined in [3] for /async requests; services should also implement it for /sync requests.

For example, if a cross match portal issues multiple requests to remote TAP services to carry out a cross-match operation, all would receive the same *RUNID*, and the service logs could later be analyzed to reconstruct the service operations initiated in response to the job.

The service **should** ensure that *RUNID* is preserved in any service logs and **should** pass on the *RUNID* value in any calls to other services.

### 2.3.9 UPLOAD

The service **should** support table upload via the *UPLOAD* parameter. The value is a list of table-name,URI pairs. Table names must be legal ADQL table names as defined in [1]. URIs maybe be simple URLs (e.g. with a URI scheme of http) or URIs (e.g. with a URI scheme of vos or param) that must be resolved to give the location of the table content. See section 2.5 for details.

### 2.3.10 Case of parameters

Parameter names **must not** be case sensitive, but parameter values **must** be case sensitive. In this document, parameter names are typically shown in uppercase for typographical clarity, not as a requirement.

### 2.3.11 Order and cardinality of parameters

Parameters in a request **may** be specified in any order.

When request parameters are duplicated with conflicting values, the response from the service is undefined. The service **may** reject the request or it **may** pick one value for the parameter. Clients **should not** repeat parameters in a request.

## *2.4 Table names*

A fully qualified table name has the form

```
[[catalog_name"."]schema_name"."]table_name
```

where *catalog_name* is the name of the DB catalogue (often the "database" name) in SQL DBMS terminology, *schema_name* is the name of the "schema" in DBMS terminology (often also called a "database"; a DBMS schema is a type of data model where the top level data model elements are tables), and *table_name* is the actual table name. All elements of the table name are optional except *table_name*. Depending upon the DBMS, "catalog" or "schema" may or may not be implemented; some DBMS implement both, others one or the other, and the simplest database systems might not implement either.

The implementation of a TAP service **must** define the table names acceptable in queries and **must** reveal these to clients through metadata queries or through VOSI-tables output, and the names **must** be identical in each of these sources. A TAP client must determine the acceptable names from one of these sources or from the cached form of the VOSI-tables output included in the service's registration.

## *2.5 Table Upload*

The service **should** implement the table upload capability. If upload is supported, the service must accept tables in VOTable format. The client specifies the name of the uploaded table; this name **must** be a legal ADQL table name with no catalog or schema (e.g. an unqualified table name). Uploaded tables **must** be referred to in queries as *TAP_UPLOAD.<tablename>*, where *<tablename>* is the specified by the user*.

Tables in the *TAP_UPLOAD* schema are transient and persist only for the lifetime of the query (although caching might be used behind the scenes) and are never visible in the TAP_SCHEMA metadata.

The column names in the transient database table are taken directly from the name attribute of the VOTable FIELD and PARAM elements. The datatypes of the transient table are determined from the FIELD and PARAM attributes as follows:

| VOTable: datatype | VOTable: arraysize | VOTable: xtype TAP_SCHEMA.columns: datatype | database column type |
|---|---|---|---|
| boolean | [1] | | Not supported |
| short | [1] | | SMALLINT |
| int | [1] | | INTEGER |
| long | [1] | | BIGINT |
| float | [1] | | REAL |
| double | [1] | | DOUBLE |
| <numeric type> | > 1 | | VARBINARY |
| char | [1] | | CHAR(1) |
| char | | | VARCHAR[3] |
| char | n* | | VARCHAR(n) |
| char | n | | CHAR(n) |
| unsignedByte | | | VARBINARY[4] |
| unsignedByte | n* | | VARBINARY(n) |
| unsignedByte | n | | BINARY(n) |
| unsignedByte | n, *, n* | adql:BLOB | BLOB |
| char | n, *, n* | adql:CLOB | CLOB |
| char | n, *, n* | adql:TIMESTAMP | TIMESTAMP |
| char | n, *, n* | adql:POINT | POINT |
| char | n, *, n* | adql:REGION | REGION |

The default mapping of data types are shown above (no arraysize or xtype). If the xtype attribute is set, this is the preferred internal datatype. If xtype is not set, then the datatype and arraysize indicate the most suitable internal datatype.

In the arraysize column above, [1] means the arraysize is not set or is set to 1, n means arraysize is set to a specific value, * means arraysize="*", and n* means arraysize="n*" (variable size up to length n). A blank means the arraysize is not set.

Binary values (unsignedByte in VOTable, BINARY, VARBINARY, or BLOB in ADQL) can be expressed as specified by the VOTable standard. By default,

---

3  This is the default internal datatype for character values. The service implementation must choose a suitable size for the VARCHAR column

4  This is the default internal datatype for binary values. The service implementation must choose a suitable size for the VARBINARY column

VOTable allows them to be written as an array of decimal numbers, e.g. 12 56 0 255 0 0 255 (one number per byte value).

For columns of type BLOB or CLOB, most database systems support reference to these columns in the select clause but not in any other part of the query. Services may use these types to indicate that columns may only be selected. For example, if service implementors want to make URL(s) available as column values in the results, but do not actually store the URL(s) in the database, they would specify a column with xtype="adql:CLOB" and the column with URL(s) could be referenced in the SELECT clause of a query, but could not be used in the WHERE clause. The service could then process the query result and insert the URL(s) or, more likely, transform a column value (an identifier) into a URL while writing the results.

TIMESTAMP values are specified using ISO8601 format without a timezone (as in   2.3.4 ) and are assumed to be in UTC. The xtype="adql:TIMESTAMP" attribute must be specified in an uploaded VOTable in order for the values to be inserted in a column of type TIMESTAMP; without the xtype, the values would be inserted into a CHAR(n) or VARCHAR column.

POINT and REGION values are specified in STC-S format (see section 6 ). The xtype="adql:POINT" attribute must be specified in an uploaded VOTable in order for the char values to be parsed and treated as POINTs (e.g. to be used with some of the ADQL region functions). For regions, the xtype="adql:REGION" attribute must be specified in an uploaded VOTable in order for the char values to be parsed and treated as REGIONs (e.g. to be used with some of the ADQL region functions).

### 2.5.1 UPLOAD

The *UPLOAD* parameter is used to reference read-only external tables via their URI, to be uploaded for use as input tables to the query.    The value of the *UPLOAD* parameter is a list of table name-URI pairs. Elements of the list are delimited by semicolon and the two parts of the pair are delimited by comma. For example:

```
UPLOAD=table_a,http://host_a/path;table_b,http://host_b/path
```

would define two input tables *table_a* and table_*b*, located at the given URIs. Services that implement *UPLOAD* **must** support *http* as a URI scheme (e.g. must support treating an *http* URI as a URL). A VOSpace URI (*vos:<something>*)  is a more generic example of a URI that requires more service-side functionality; support for the *vos* scheme is optional.

### 2.5.2 Inline Table Upload

To upload a table inline, the caller must specify the UPLOAD parameter (as above) using a special URI scheme "param". This scheme indicates that the value after the colon will be the name of the inline content. The content type used is *multipart/form-data*, using a "file" type input element. The "name" attribute must match that used in the UPLOAD parameter.

For example, in the POST data we might have this parameter:

```
UPLOAD=table_c,param:table1
```

and this content:

```
Content-Type: multipart/form-data; boundary=AaB03
[...]
--AaB03x
Content-disposition: form-data; name="table1"; filename="table1.xml"
Content-type: application/x-votable+xml
[...]
--AaB03x
[...]
```

The uploaded table would be referenced in queries as *TAP_UPLOAD.table_c* (the table name in the UPLOAD parameter). Services that implement table upload **must** support the *param* scheme for inline uploads.

In principle, any number of tables can be uploaded using the UPLOAD parameter and any combination of URI schemes supported by the service as long as they are assigned unique table names within the query. Services may limit the size and number of uploaded tables; if the service refuses to accept the entire table it **must** respond with an error as described in 2.7.3 .

## *2.6 Metadata and TAP_SCHEMA*

There are several approaches to getting metadata for a given TAP service. All TAP services **must** support a set of tables in a schema named *TAP_SCHEMA* that describe the tables and columns included in the service. In addition to the *TAP_SCHEMA*, there are two other ways to get metadata from a TAP service. First, the VOSI tables resource provides metadata on all tables and columns; this resource is described in 2.2.5 . The VOSI tables resource provides the same metadata as the *TAP_SCHEMA* but in a rigorously controlled format; the information in the *TAP_SCHEMA* is equivalent to that defined by the VODataService [7]. Second, the client may specify a query of one or more tables setting the *MAXREC* parameter to 0 so that only the metadata regarding the requested fields is returned. Use of *MAXREC* is described in 2.3.7 .

The *TAP_SCHEMA* provides access to table, column, and join key metadata through the TAP query mechanisms themselves. Users can discover tables or columns that meet their specific criteria by querying the tables described below. The service may enhance the *TAP_SCHEMA* with additional metadata where that seems appropriate; since it is self-describing, the *TAP_SCHEMA* may be queried to determine if any extended schema metadata is defined by the service. Services must provide these tables and make them accessible by all supported query mechanisms.

The qualified names in the tables of the TAP schema **must** follow the rules defined in section 2.4. The names **must** be stated in a form that is acceptable as an operand of a query.

All columns in the TAP_SCHEMA tables are of type VARCHAR except for size, principal, indexed, and std (in Columns) which are INTEGER values.

Implementors are permitted to include additional tables in the TAP_SCHEMA to describe additional aspects of their service not covered by this specification. Implementors may also include additional columns in the standard tables described below. For example, one could include a column with a timestamp saying when metadata values were was last modified.

### 2.6.1 Schemas

The table TAP_SCHEMA.schemas **must** contain the following columns:

| Column name | datatype | |
|---|---|---|
| schema_name | varchar | schema name, possibly qualified |
| description | varchar | brief description of schema |
| utype | varchar | UTYPE if schema corresponds to a data model |

The schema_name values **must** be unique and may be qualified by the catalog name or not depending on the implementation requirements. The fully qualified schema name is defined by the ADQL language and follows the pattern *[catalog.]schema.* The schema metadata are included for reference and are not used directly to construct queries.

### 2.6.2 Tables

The table TAP_SCHEMA.tables **must** contain the following columns:

| Column name | datatype | |
|---|---|---|
| schema_name | varchar | the schema name from TAP_SCHEMA.schemas |
| table_name | varchar | table name as it should be used in queries |
| table_type | varchar | one of: table, view |
| description | varchar | brief description of table |
| utype | varchar | UTYPE if table corresponds to a data model |

The table_name values **must** be unique. The value of the table_name should be the string that is recommended for use in querying the table; it may or may not be qualified by schema and catalog name(s) depending on the implementation requirements. The fully qualified table name is defined by the ADQL language and follows the pattern *[[catalog.]schema.]table*.

### 2.6.3 Columns

The table TAP_SCHEMA.columns must contain the following columns:

| Column name | datatype | |
|---|---|---|
| table_name | varchar | table name from TAP_SCHEMA.tables |
| column_name | varchar | column name |
| description | varchar | brief description of column |
| unit | varchar | unit in VO standard format[5] |
| ucd | varchar | UCD of column if any |
| utype | varchar | UTYPE of column if any |
| datatype | varchar | ADQL datatype as in section  2.5 |
| size | integer | length of variable length datatypes[6] |
| principal | integer | a principal column; 1 means true, 0 means false |
| indexed | integer | an indexed column; 1 means true, 0 means false |
| std | integer | a standard column; 1 means true, 0 means false |

The table_name,column_name (pair) values **must** be unique.

Data types and how they map to VOTable datatypes are described in section  2.5 above. The "size" gives the length of variable length datatypes, for example varchar(256); this size does not map to the VOTable arraysize attribute when the latter specifies the size and shape of a multi-dimensional array. The "principal" flag indicates that the column is considered a core part the content; clients can use this hint to make the principal column(s) visible, for example by selecting them by default in generating an ADQL query. In cases where the services selects the columns to return (such as PQL without a SELECT parameter), the principal column indicates those columns that are returned by default. The "indexed" flag indicates that the column is indexed, potentially making queries run much faster if this column is used in a constraint. The "*std*" is included for compatibility with the registry, which uses this value to indicate that a given column is defined by some standard, as opposed to a custom column defined by a particular service.

### 2.6.4 Foreign Keys

The table TAP_SCHEMA.keys **must** contain the following columns to describe foreign key relations between tables:

| **Column name** | **datatype** | |
|---|---|---|
| key_id | varchar | unique key identifier |
| from_table | varchar | fully qualified table name |
| target_table | varchar | fully qualified table name |
| description | varchar | description of this key |
| utype | varchar | utype of this key |

The key_id values are unique and used only to join with the TAP_SCHEMA.key_columns table below. There may be one or more rows with different key_id values and a pair of tables to denote one or more ways to join the tables.

---

5  There is no actual standard, but by convention IVOA services and applications use SI units and FITS units; units used here should conform to those allowed in VOTable [9].

6  Variable length datatypes include CHAR, VARCHAR, BINARY, and VARBINARY. The size should be NULL for fixed size datatypes (numeric, timestamp) and may be NULL for arbitrary-sized columns (CLOB, BLOB).

The table TAP_SCHEMA.key_columns **must** contain the following columns to describe the columns that make up a foreign key :

| Column name | datatype | |
|---|---|---|
| key_id | varchar | key identifier from the TAP_SCHEMA.keys |
| from_column | varchar | key column name in the <from_table> |
| target_column | varchar | key column in the <target_table> |

There may be one or more rows with a specific key_id to denote single or multi-column keys.

A TAP service **must** provide the tables listed above and **may** provide other tables in the *TAP_SCHEMA* namespace.

## 2.7 Access to and Representations of Results

### 2.7.1 Data and metadata queries

The result of a data query or a metadata query depends on the query language used and may be one or more tables in one or more resources. Unsupportable combinations of query result and FORMAT (e.g. queries that produce multiple tables and an inherently single-table format like CSV) will cause the request to fail. Currently, an ADQL query result **must** be a single table (in a single file).

This table **must** be encoded in the output format specified by the FORMAT parameter of the query. See section  2.3.6  for required, optional and default formats. VOTable is the default format and VOTable support is mandatory.

The output table **must** include the same number and order of columns as specified in the SELECT clause of the query. For VOTable output, the name attribute of FIELD elements **must** be the same as the column names (or aliases if specified in the query) from the query and the datatype, arraysize, and xtype attributes of FIELD elements **must** be set using the mapping specified in section  2.5 . The xtype attribute in the output **must** match the datatype for the column in the TAP_SCHEMA.

VOTable structure follows the rules in section  2.9  and **must** be returned with an allowed VOTable MIME type (*application/x-votable+xml* or *text/xml)*. If the FORMAT parameter (see  2.3.6 ) of the request specified a specific VOTable MIME type, the requested MIME type **must** be used in the HTTP response.

CSV formatted data **should** represent the output table with one row of text per table row, with the table column values rendered as text and separated by commas. If a column value contains a comma the entire column value **should** be enclosed in double quotes.  Text lines may be arbitrarily long.  The first data row **should** give the column name as the data value.   CSV data **must** be returned with a MIME type of *text/csv*; if the optional header line (with column names) is included, the MIME type must be *text/csv;header=present*. Full details of CSV format are defined in RFC 4180 [14].

TSV formatted data **should** represent the output table with one row of text per table row, with the table column values rendered as text and separated by the TAB character. TSV data **must** be returned with a MIME type of *text/tab-separated-values* [15]. Column values may not contain the TAB character.

## 2.7.2 VOSI

Representations of VOSI outputs (capabilities, availability, table metadata) **must** be as defined in the VOSI standard [6].

The representation of table metadata **must** include all tables in the service's tableset. VOSI's representation of table metadata is specified in *VODataService* [7].

The VOSI standard specifies that the capability metadata is encoded as an XML document which lists each of the service's capabilities as a <capability> element. The type of this element (which defines the contents) is {http://www.ivoa.net/xml/VOResource/v1.0}Capability from the VOResource XML standard [8].

In addition, the capabilities output must also comply with the following requirements:

· the returned document **must** include one <capability> element that describes the service's support for the TAP protocol

· this <capability> element **must** have its "standardID" attribute set to "ivo://ivoa.net/std/TAP"

· this capability element **must** include at least one "<interface>" element with its "role" attribute set to "std",

· this "<interface>" element **must** contain a child "<accessURL>" element with the attribute "use" set to "base" which contains the root web resource for the service as defined in section  2.2 .

  *Note: VO registries recognize a service's support for a standard protocol through this capability description. In particular, a different standard Capability sub-type is used for each standard protocol to provide capability metadata that is specific to that protocol. At the time of this writing, a Capability sub-type for TAP has not yet been defined. Thus for compliance with this standard, any legal Capability description that meets the above restrictions is sufficient. However, once a VOResource extension for TAP is standardized, it is strongly recommended that TAP services emit its capabilities using that the Capability sub-type specialized for TAP.*

For example, the returned capabilities document for a service supporting   TAP might look as follows:

```xml
<?xml version="1.0" encoding="UTF-8"?>
<vosi:capabilities xmlns=""
    xmlns:vosi="http://www.ivoa.net/xml/VOSI/v1.0"
    xmlns:vs="http://www.ivoa.net/xml/VODataService/v1.0"
    xmlns:xsi="http://www.w3.org/2001/XMLSchema-instance"
    xsi:schemaLocation="http://www.ivoa.net/xml/VOSI/v1.0
                        http://www.ivoa.net/xml/VOSI/v1.0
             http://www.ivoa.net/xml/VODataService/v1.0
             http://www.ivoa.net/xml/VODataService/v1.0">
  <vosi:capability standardID="ivo://ivoa.net/std/TAP">
    <interface xsi:type="vs:ParamHTTP" role="std">
      <accessURL use="base"> http://myarchive.net/myTAP </accessURL>
    </interface>
  </vosi:capability>
  <vosi:capability standardID="ivo://ivoa.net/std/VOSI#capabilities">
    <interface xsi:type="vs:ParamHTTP">
      <accessURL use="full">
        http://myarchive.net/myTAP/capabilities </accessURL>
    </interface>
  </vosi:capability>
  <vosi:capability standardID="ivo://ivoa.net/std/VOSI#availability">
    <interface xsi:type="vs:ParamHTTP">
      <accessURL use="full">
        http://myarchive.net/myTAP/availability
      </accessURL>
    </interface>
  </vosi:capability>
  <vosi:capability standardID="ivo://ivoa.net/std/VOSI#tables">
    <interface xsi:type="vs:ParamHTTP">
      <accessURL use="full">
        http://myarchive.net/myTAP/tables </accessURL>
    </interface>
  </vosi:capability>
</vosi:capabilities>
```

### 2.7.3 Errors

If the service detects an exceptional condition, it **must** return an error document with an appropriate HTTP-status code. TAP distinguishes three classes of exceptions.

- Errors in the use of the HTTP protocol.
- Errors in the use of the TAP protocol, including both invalid requests and failure of the service to complete valid requests.

Error documents for HTTP-level errors are not specified in the TAP protocol. Responses to these errors are typically generated by service containers and cannot be controlled by TAP implementations. There are several cases where a TAP service could return an HTTP error. First, the /async endpoint could return a 404 (not found) error if the client accesses a job within the UWS joblist that does not exist. Second, access to a resource could result in an HTTP 401 (not authorized) error if authentication is required or an HTTP 403 (forbidden) error if the client is not allowed to access the resource.

Error documents for TAP errors **must** be VOTable documents;  any result-format specified in the request is ignored. If the error document is being retrieved from the /async/<jobid>/error resource (specified by UWS) after an asynchronous query, the HTTP status code should be 200. If the error document is being returned directly after a synchronous query, the service may use an appropriate HTTP status code, including 200 (successfully returning a response to the request) and various 4xx and 5xx values. The exception condition **must** be described to the client using a status code in the VOTable header.  Section  2.9 specifies the use of VOTable for error documents in TAP services.

### 2.7.4 Overflows

If a query is executed by a TAP service, the number of rows in the table of results may exceed a limit requested by the user (using the MAXREC parameter) or a limit set by the service implementation (the default or maximum value of MAXREC). In these cases, the query is said to have 'overflowed'. Typically, a TAP service will not detect an overflow until some part of the table of results has been sent to the client.

If an overflow occurs, the TAP service **must** produce a table of results that is valid, in the required output format, and which contains all the results up to the point of overflow. Since an output overflow is not an error condition, the MIME type of the output **must** be the same as for any successful query and the HTTP status-code **must** be as for a successful, complete query.

If the output format is VOTable, section  2.9.1  describes the method by which the overflow is reported. No method of reporting an overflow is defined for formats other than VOTable.

## 2.8 Versioning of the TAP protocol

The TAP protocol provides explicitly for versioning of the interface in order to support version negotiation between a client and a service where one or both parties support more than one version of the protocol. The TAP version refers only to the TAP protocol; query languages are versioned separately and TAP and ADQL versions may differ.

Version numbers follow IVOA document conventions [17].

### 2.8.1 Appearance in requests and in service metadata

The version number may appear in at least three places: in the service metadata, as a parameter in client requests to a server, and in the query response. The version number used in a client's request of a particular server must be equal to a version number which that server has declared it supports (except during negotiation, as described below). A server may support several versions, whose values clients may discover according to the negotiation rules.

### 2.8.2 Version number negotiation

If a TAP client does not specify the version number in a request, the server assumes the highest standard version supported by the service, and no explicit version checking takes place.   If the client specifies an explicit version number, and this does not match a version available from the service, the service returns a version number mismatch error as described in   2.9.2 . The client can determine what versions of the protocol the service supports by a prior call to VOSI-capabilities or via a registry query.

## 2.9 Use of VOTable

VOTable is a general format. TAP requires that it be used in a particular way.

The result VOTable document **must** comply with VOTable v1.2 or greater [9]. For columns containing coordinate values, the coordinate system metadata should be provided as described in [13].

The VOTable **must** contain a *RESOURCE* element identified with the attribute *type="results"*, containing a single *TABLE* element with the results of the query. Additional *RESOURCE* elements may be present, but the usage of any such elements is not defined here and TAP clients **should not** depend upon them.

### 2.9.1 INFO elements

The *RESOURCE* element **must** contain, before the *TABLE* element, an *INFO* element with attribute *name* = *"QUERY_STATUS"*. The *value* attribute **must** contain one of the following values:

- "OK", meaning that the query executed successfully and a result table is included in the resource

- "ERROR", meaning that an error was detected at the level of the TAP protocol or the query failed to execute

The content of the INFO element conveying the status **should** be a message suitable for display to the user describing the status.

**Example:**

<INFO name="QUERY_STATUS" value="OK"/>

**Example:**

<INFO name="QUERY_STATUS" value="OK">Successful query</INFO>

**Example:**

```
<INFO name="QUERY_STATUS" value="ERROR">
  value out of range in POS=45,91
</INFO>
```

Additional *INFO* elements **may** be provided, e.g., to echo the input parameters back to the client in the query response (a useful feature for debugging or to self-document the query response), but clients **should not** depend on these.

**Example:**

```
<RESOURCE type="results">
<INFO name="QUERY_STATUS" value="ERROR">
    unrecognized operation
</INFO>
<INFO name="SPECIFICATION" value="TAP"/>
<INFO name="VERSION" value="1.0"/>
<INFO name="REQUEST" value="doQuery"/>
<INFO name="baseUrl" value="http://webtest.aoc.nrao.edu/ivoa-dal"/>
<INFO name="serviceVersion" value="1.0"/
...
</RESOURCE>
```

If an overflow occurs (result exceeds MAXREC), the service must close the table and append another INFO element to the RESOURCE (after the TABLE) with *name="QUERY_STATUS"* and the value="*OVERFLOW*".

**Example:**

```
<RESOURCE type="results">
<INFO name="QUERY_STATUS" value="OK"/>
...
<TABLE>...</TABLE>
<INFO name="QUERY_STATUS" value="OVERFLOW"/>
</RESOURCE>
```

In the above example, the TABLE should have exactly MAXREC rows.

If an error occurs while writing the rows of the VOTable, the service must close the table and append another INFO element to the RESOURCE, after the TABLE, with *name="QUERY_STATUS"* and the value="ERROR*".

**Example:**

```
<RESOURCE type="results">
<INFO name="QUERY_STATUS" value="OK"/>
...
<TABLE>...</TABLE>
<INFO name="QUERY_STATUS" value="ERROR" />
</RESOURCE>
```

The content of these trailing INFO elements is optional and intended for users; client software **should not** depend on it.

Thus, one INFO element with *name="QUERY_STATUS"* and *value="OK"* or *value="ERROR"* **must** be included before the TABLE. If the TABLE does not contain the entire query result, one INFO element with *value="OVERFLOW"* or *value="ERROR"* **must** be included after the table.

## 2.9.2 Version Mismatch Errors

Errors due to version mismatch from either the VERSION parameter (TAP version) or specific version used in the LANG parameter (query language version) are specified using an INFO element with *name="QUERY_STATUS"* and *value="ERROR"* as described above.

# 3 Service Registration (normative)

Publication of a service to the VO requires that it be registered with an IVOA registry, including describing the identity and capabilities of the service.

The resource document for a TAP service instance **must** be structured according to *VOResource* [8] using the sub-type *CatalogService* as defined in *VODataService* [7].

The resource document **must** include a *capability* element denoting the TAP interface and functions. This element must contain the URL for the root web resource (as defined in section 2.2 ). Clients would add to this URL /sync or /async as appropriate.

The resource document **must** contain capability elements for the VOSI-capabilities, VOSI-availability and VOSI-tables outputs. These **must** be formatted as in the VOSI standard [6].

The resource document should include the table metadata, except where the database-schema of the archive changes frequently. Where table metadata are provided, they **must** be represented as XML elements drawn from *VODataService* [7].

# 4 Extended capabilities (normative)

The TAP service allows for optional extended capabilities and operations. Extensions may be defined within an information community when needed for additional functionality or specialization. A generic client **must** not be required or expected to make use of such extensions. Extended capabilities or operations **must** be defined by the service metadata. Extended capabilities provide additional metadata about the service, and may or may not enable optional new parameters to be included in operation requests. Extended operations may allow additional operations to be defined.

A server **must** produce a valid response to the operations defined in this document, even if parameters used by extended capabilities are missing or malformed (i.e. the server **must** supply a default value for any extended capabilities it defines), or if parameters are supplied that are not known to the server.

Service providers **must** choose extension names with care to avoid conflicting with standard metadata fields, parameters and operations.

# 5 Use of UWS (informative)

The UWS pattern is specified in [3] and its application to TAP in section 2.2.2 . This section explains the exchange of messages between a TAP client and service when using UWS to run an asynchronous query.

Consider a TAP service at *http://example.com/tap*. TAP mandates that the asynchronous requests be directed to *http://example.com/tap/async*. This URL points to the list of 'jobs'; i.e. the list of queries currently or recently executed.

## 5.1 Creating a Query

To create a new query, the client POSTs a request to the job list:

```
HTTP POST http://example.com/tap/async
REQUEST=doQuery
LANG=ADQL
QUERY=SELECT TOP 100 * FROM foo
```

The service then creates a job and assigns that job a name and a URL based on the name. Suppose that the name is *42*, then the URL will be *http://example.com/tap/async/42* because the jobs are always children of the job list. While the job is in the PENDING phase, additional parameters may be specified by additional POSTs to the job resource, for example:

```
HTTP POST http://example.com/tap/async/42
UPLOAD=mytable,http://a.b.c/mytable.xml
```

After each such POST, the service issues an HTTP redirection to the job's URL, where the modified state may be accessed:

```
HTTP status 303 'See other'
Location: http://example.com/tap/async/42
```

All TAP-specific parameters are stored using the paramList mechanism of UWS and are included in the XML representation of the job:

```
HTTP GET http://example.com/tap/async/42
```

or directly from the parameters resource:

```
HTTP GET http://example.com/tap /async/42/parameters
```

Individual parameters cannot be accessed as separate web resources.

The UWS pattern requires the following resources to describe and control the job:

```
http://example.com/tap/async/42/phase
http://example.com/tap/async/42/quote
http://example.com/tap/async/42/executionduration
http://example.com/tap/async/42/destruction
http://example.com/tap/async/42/results
http://example.com/tap/async/42/error
```

The quote resource specifies the predicted completion time for the job (query), assuming it is started immediately. In practice, it is very hard to estimate the time a query will take; for TAP services it is recommended that this be set to the current time plus the maximum amount of time the query will be allowed to run (see termination below).

The termination resource specifies the amount of time (in seconds) the job (query) will be allowed to run before being aborted by the service. The termination time is set by the service and can be read from the job or directly from the termination resource:

```
HTTP GET http://example.com/tap/async/42/executionduration
```

The service may allow the client to change the termination:

```
HTTP POST http://example.com/tap/async/42/executionduration
TERMINATION=600
```

The destruction resource specifies when the service will destroy the job. The service is only required to keep a job for a finite period of time, after which it may destroy the job, including the result. After this time, the client will receive an HTTP 404 'not found' status if it tries to get any information about the job. The destruction time of the job is chosen by the service and the client can read it from the job or directly from the destruction resource:

```
HTTP GET http://example.com/tap/async/42/destruction
```

The service may allow the client to change the destruction time:

```
HTTP POST http://example.com/tap/async/42/destruction
DESTRUCTION=2008-11-11T11:11:11Z
```

## 5.2 Running the Query

The *phase* URL shows the progress of the job. When the job is created by the service it will normally be set to *PENDING*, but might be set to *ERROR* if the service has rejected the job. If the phase is *ERROR*, then the *error* URL should lead to a an error document explaining the problem. If the phase is *PENDING*, then the client needs to commit the job for execution.

The client **runs** the job by posting to the phase URL:

```
HTTP POST http://example.com/tap /async/42/phase
PHASE=RUN
```

The service replies with a redirection to the job URL

```
HTTP status 303 'see other'
Location: http://example.com/tap /async/42
```

The phase will now have changed to either *QUEUED* or *EXECUTING*, depending on the service implementation. The client **tracks** the execution by polling the phase URL:

```
HTTP GET http://example.com/tap/async/42/phase
```

A job in the QUEUED or EXECUTING phase may be **aborted** by posting to the phase URL:

```
HTTP POST http://example.com/tap/async/42/phase
PHASE=ABORT
```

When the query is complete, the phase changes to *COMPLETED*. The client then retrieves the result from the results list:

```
HTTP GET http://example.com/tap/async/42/results/result
```

The client knows that the table of results is at the URL /result relative to the results list because the TAP protocol requires this naming. A generic UWS client could find the name of the result and retrieve it by examining either the job description:

```
HTTP GET http://example.com/tap/async/42
```

or by looking specifically at the result list:

```
HTTP GET http://example.com/tap/async/42/results
```

If the service cannot run the query, then the final phase is *ERROR* and there is no table of results. In this case, the client should expect an HTTP 404 'not found' status if it tries to retrieve the result. The client should look instead at the error resource to find out what went wrong:

```
HTTP GET http://example.com/tap/async/42/error
```

If the job was aborted (by the client or the service), the final phase will be ABORTED and there is no table or results. As with errors, the client should look at the error resource to find out what went wrong.

The basic sequence can be executed from a web browser or from a shell script using the *curl* utility:

```
curl -d 'REQUEST=doQuery&LANG=PQL&POS=12,34&SIZE=0.5&FROM=foo' \
        http://example.com/tap/async
  [read Location header from curl output]
curl -d 'PHASE=RUN' http://example.com/tap/async/42
curl http://example.com/tap/async/42/phase
[repeat until phase is COMPLETED]
curl http://example.com/tap/42/results/result
```

# 6 Use of STC-S in TAP (informative)

In the absence of an accepted standard for the string serialisation of geometry values, we recommend that TAP service and client implementers make use of the simple scheme described here. This scheme is based on the STC-S 1.30 IVOA Note and values for string constants from the STC 1.33 Recommendation. We also provide a simple extrapolation for compound regions. We have limited STC-S usage to include the set of geometry values that can be constructed by the other ADQL geometry functions (POINT, CIRCLE, BOX, POLYGON); the same values can thus be used everywhere geometry values need to be represented.

> *WARNING: The syntax and usage here will be deprecated in the future once a standard syntax for geometry is recommended by the IVOA. At that time, a revised TAP specification will be developed to explicitly specify the usage of STC within TAP. Until that time, the text in this section is informative only, but we suggest that services attempt to implement some or all of these suggestions if applicable.*

STC-S strings can be used in three ways with TAP:

- as the argument to the ADQL REGION function

- as values in a query result when the column has datatype (in the TAP_SCHEMA.columns table) and xtype (in the VOTable FIELD) of adql:POINT or adql:REGION

- as values in an uploaded table when the FIELD describing the column has xtype adql:POINT or adql:REGION

## 6.1 Simple STC-S BNF

```
<region> ::= <position> | <circle> | <box> | <polygon> | <union> |
<intersection> | <not>

<position> ::= POSITION <coordsys> <coordpair>

<circle> ::= CIRCLE <coordsys> <coordpair> <radius>

<box> ::= BOX <coordsys> <coordpair> <width> <height>

<polygon> ::= POLYGON <coordsys> <coordpair> <coordpair> <coordpair>
[<coordpair> ...]

<union> ::= UNION <coordsys> ( <region> <region> [<region> ... ] )

<intersection> ::= INTERSECTION [<coordsys>] ( <region> <region>
[<region> ... ] )

<not> ::= NOT ( <region> )

<coordsys> ::= [<frame>] [<refpos>] [<flavor>]

<coordpair> ::= <coord1> < coord2>

<coord1> ::= <numeric value>
```

```
<coord2> ::= <numeric value>
<width> ::= <numeric value>
<height> ::= <numeric value>

<frame> ::= ECLIPTIC | FK4 | FK5 | GALACTIC | ICRS |  UNKNOWNFRAME
<refpos> ::= BARYCENTER | GEOCENTER | HELIOCENTER | LSR | TOPOCENTER
| RELOCATABLE | UNKNOWNREFPOS
<flavor> ::= CARTESIAN2 |  CARTESIAN3 | SPHERICAL2
```

Although shown above in upper case for clarity, all string constants are case-insensitive. Thus, CIRCLE, circle, Circle, and CiRcLe are all equivalent. The <numeric value> is a standard number; it may be expressed in scientific notation.

The default value for frame is UnknownFrame. For FK4 the assumed equinox is B1950 and for FK5 it is J2000; there is no provision to specify any other equinox.

The default reference position is UnknownRefPos.

The default coordinate flavor is SPHERICAL2. For consistency with other ADQL geometry functions, the units for numeric values in spherical coordinates are always degrees. For Cartesian coordinates, the units are arbitrary and assumed to be local units; the user must examine the appropriate column metadata and express values in the specified units.

The <coordsys> value is composed of the optional frame, an optional reference position and optional coordinate flavor; the entire coordinate system may be unspecified and take default values. This value may also be used as the first argument to the other ADQL geometry functions, in which case the zero-length string, a string with only spaces, and the NULL value are equivalent to UnknownFrame UnknownRefPos Spherical2 in STC-S. TAP services should assume that values in unknown frames (or with unknown reference positions) are in the same frame (reference position) as the column(s) being compared. Values in Cartesian coordinate flavors must always use UnknownFrame and UnknownRefPos (or simply leave them out since this is the default).

The full STC model specifies operators for combining geometry values into compound regions. Here we recommend union, intersection, and not operators because of the simple relationship they have with OR and AND in query languages like ADQL. In addition, UNION is needed to describe some geometry values in use in astronomy: regions described by disconnected simple regions. See  6.6  below for an example of how UNION could apply to a real-world situation and how OR can be used interchangeably. When specifying an operator, the <coordsys> should be specified with the outermost operator and not with each <region> included (e.g. in the most compact form and with only one <coordsys> specified). The above BNF does not impose the restriction, but we still recommend that STC-S expressions be written this way for simplicity.

## 6.2 Example 1: Circle on the Sky in ICRS Coordinates

A circle centered at RA,DEC = 10,20 with radius 0.5 degrees is expressed in STC-S as

```
Circle ICRS GEOCENTER 10 20 0.5
```

When used in a query, it would be used as

```
Region('Circle ICRS GEOCENTER 10 20 0.5')
```

and is equivalent to the ADQL function

```
Circle('ICRS GEOCENTER', 10, 20, 0.5).
```

All numeric values are in degrees since the circle uses the default SPHERICAL2 coordinate flavor.

## 6.3 Example 2: Position in Galactic Coordinates (l,b)

A position of l,b = 10,20 is expressed in

```
STC-S as Position GALACTIC 10 20
```

When used in a query, it would be used as

```
Region(' Position GALACTIC 10 20')
```

and is equivalent to the ADQL function

```
Point('GALACTIC', 10, 20).
```

All numeric values are in degrees since the position uses the default SPHERICAL2 coordinate flavor. The reference position is unknown.

## 6.4 Example 3: Box in Cartesian Coordinates

A box from 2,2 to 4,4 (e.g. centered at 3,3) in a two-dimensional Cartesian coordinate system is expressed in STC-S as

```
Box CARTESIAN2 3 3 2 2
```

When used in a query it would be used as

```
Region('Box CARTESIAN2 3 3 2 2')
```

and is equivalent to the ADQL function

```
Box('CARTESIAN2', 3, 3, 2, 2).
```

In this case, the TAP service would interpret the values as being in the same coordinate system (frame and reference position) as is used internally since the values are unknown.

## 6.5 Example 4: Box on the Sky in Local Coordinates

A box with great circle sides of length 2 degrees and centered at 180,0 is expressed in STC-S as

```
Box 180 0 2 2
```

When used in a query it would be used as

```
Region('Box 180 0 2 2')
```

and is equivalent to either of these uses of the ADQL function

```
Box(' ', 180, 0, 2, 2)
Box(NULL, 180, 0, 2, 2).
```

The default coordinate flavor is SPHERICAL2 so the coordinate values are longitude and latitude in degrees. The frame is UnknownFrame and the reference position is UnknownRefPos, so the TAP service should treat the box as being in the relevant local coordinate system.

### 6.6 Example 5: A Region made up of Two Unconnected Polygons

An observation from a mosaic camera with a large gap between detectors can be described as the union of two polygons. In STC-S this is

```
Union ICRS (Polygon 1 4 2 4 2 5 1 5 Polygon 3 4 4 4 4 5 3 5)
```

where the two polygons are 1x1 degrees and the gap between is 1 degree (from a missing chip or bad data, for example). The longitude increases to the right and the area included in the polygons is to the left of the edges (left-hand-rule). Such a value could appear in the query result from a TAP service where observations are the content being queried. This footprint could be used in a query:

```
CONTAINS(someTable.position, REGION('Union ICRS (Polygon 1 4 2 4 2 5
1 5 Polygon 3 3 4 4 4 4 5 3 5 )')) = 1
```

The equivalent query condition using the ADQL POLYGON function would be:

```
( CONTAINS( someTable.position, Polygon('ICRS',1,4,2,4,2,5,1,5)) = 1
OR
CONTAINS( someTable.position, Polygon('ICRS',3,4,4,4,4,5,3,5)) = 1 )
```

where someTable.position is a column with datatype adql:POINT (in the TAP_SCHEMA). This example also shows the logical equivalence of UNION and OR.

### 6.7 Example 6: A Polygon with a Hole

A polygon with a hole is described in STC-S by the intersection of the polygon with the negation of the polygon describing the (triangular) hole: Intersection ( Polygon 1 1 4 1 4 4 1 4 Not ( Polygon 2 2 3 2 2 3 ) ). Here, the vertices of both polygons are given in counter-clockwise order so the contained region is inside (small). For the second region of the intersection (the NOT) one could also reverse the order of the vertices instead of using the NOT operator to specify that the contained region is outside rather than inside the triangle: Intersection ( Polygon 1 1 4 1 4 4 1 4 Polygon 2 2 2 3 3 2 ) ). Thus, the NOT operator is equivalent to reversing the order of the vertices.

Note that the coordinate frame (ICRS) is specified once outside the brackets for the union; it could also be specified inside the brackets for each polygon as described above (after the Polygon token and before the coordinate values). We have simply shown the more compact representation. In principle, one could specify the union of geometry values with different coordinate frames, but this would require coordinate transformation by the recipient of such a value (see below) so we do not recommend such usage.

## 6.8 Use of Dynamic STC-S Values with the ADQL REGION Function

The ADQL REGION function takes a single string argument and it is syntactically legal to generate the string dynamically using ADQL string concatenation (the || operator) with column references. We recommend that TAP services do not support the use of dynamic STC-S values in queries due to the implementation complexity, poor performance, and better alternatives (the other ADQL geometry functions). TAP services that detect such use should return an error.

## 6.9 Coordinate Transformations

If a geometry value is specified with a coordinate system (frame and reference position) that are different from the one used internally by the TAP service, the service should transform the value(s) to the local coordinate system and use the transformed values to execute the query. We recommend that services implement the common transformations between ICRS, FK4, FK5, and GALACTIC. However, if the service cannot perform a transformation, the query should fail with a suitable error message so users will know to perform the transformation themselves.

If the column metadata, obtained via querying the TAP_SCHEMA or by examining the content of the VOSI tables resource, indicates the coordinate system(s) used by geometry columns, we recommend that users express coordinate values in the local system as in  6.5  above. Note that the description of columns in TAP_SCHEMA does not include coordinate system metadata, but individual TAP services may extend it by including additional columns in TAP_SCHEMA.columns that describe the coordinate frame, reference position, and flavor.

Geometric values with no fixed reference position (e.g. theoretical simulations) should use RELOCATABLE for the reference position.

Finally, we recommend that TAP services use ICRS for sky coordinates where possible so that the treatment of unknown frame and reference position will usually be sensible.

# 7 VOSpace Integration (informative)

This version of TAP provides limited VOSpace integration, although better support for VOSpace is planned for a later version following further implementation experience. In this version, one may specify an upload table using a URI to a table stored in a VOSpace, e.g.:

```
HTTP POST http://example.com/tap/async/42
UPLOAD=mytable,vos://space/path/votable.xml
```

The service would resolve the URI, contact the VOSpace, retrieve the table, and make it visible to the query as TAP_UPLOAD.mytable.

A future version of TAP may specify additional use and more integration with VOSpace.

# 8  Use of HTTP (informative)

A TAP service is a web service and TAP implementations are constrained by the general rules for use of HTTP, which are contained in IETF RFC documents. This section collates some of the requirements. For authoritative specifications, please refer to the original RFCs.

## 8.1 General HTTP request rules

### 8.1.1 Introduction

This document defines the implementation of the TAP service on a distributed computing platform (DCP) comprising Internet hosts that support the Hypertext Transfer Protocol (HTTP) (see IETF RFC 2616 [11]). Thus, the Online Resource of each operation supported by a server is an HTTP Uniform Resource Locator (URL).  The URL may be different for each operation, or the same, at the discretion of the service provider.  URLs conform to the description in IETF RFC 2616 (section 3.2.2 "HTTP URL") but is otherwise implementation-dependent; only the query portion comprising the service request itself is defined by this document.

While the TAP protocol currently only supports HTTP as the DCP for general Parameterised operations, data access references are more general and may use other  internet protocols, e.g., FTP, or potentially grid protocols.

HTTP supports two primary request methods: GET and POST.  One or both of these methods may be offered by a server, and the use of the Online Resource URL differs in each case.  Support for the GET method is mandatory; support for the POST method is optional except where required for a service operation to function, e.g., uploading a large quantity of data inline in a query, or when issuing a request to the service which changes the server state.

### 8.1.2 Reserved characters in HTTP GET URLs

The URL specification (IETF RFC 2396 [5]) reserves particular characters as significant and requires that these be escaped when they might conflict with their defined usage.  This document explicitly reserves several of those characters for use in the query portion of TAP requests. When the characters "?", "&", "=", "," (comma), "/", and ";" appear in one of the roles defined in Table 1, they appear literally in the URL. When those characters appear elsewhere (for example, in the value of a parameter), they should be encoded as defined in IETF RFC 2396. The server is responsible for decoding any character escaped in this manner.

Table 1 — Reserved characters in TAP query string

| Character | Reserved usage |
|---|---|
| ? | Separator indicating start of query string. |
| & | Separator between parameters in query string. |
| = | Separator between name and value of parameter. |
| ,/; | Separator between individual values in list-oriented parameters |

In particular, if any parameter value contains the character "#" (for example in a dataset identifier) it must be URL encoded to be legally included in a URL.

## 8.1.3 HTTP GET

A URL prefix is defined in accordance with IETF RFC 2396 [5] as a string including, in order, the scheme ("http" or "https"), Internet Protocol hostname or numeric address, optional port number, path, mandatory question mark "?", and optional string comprising one or more server-specific parameters ending in an ampersand "&". The prefix defines the network address to which request messages are to be sent for a particular operation on a particular server. Each operation may have a different prefix. Each prefix is entirely at the discretion of the service provider.

This document defines how to construct a query part that is appended to the URL prefix in order to form a complete request message. Every TAP operation has several mandatory or optional request parameters. Each parameter has a defined name . Each parameter may have one or more legal values, which are either defined by this document or are selected by the client based on service metadata. To formulate the query part of the URL, a client appends the mandatory request parameters, and any desired optional parameters, as name/value pairs in the form "name=value&" (parameter name, equals sign, parameter value, ampersand). The "&" is a separator between name/value pairs, and is therefore optional after the last pair in the request string.

When the HTTP GET method is used, the client-constructed query part is appended to the URL prefix defined by the server, and the resulting complete URL is invoked as defined by HTTP (IETF RFC 2616).

Table 2 summarizes the components of an operation request URL when HTTP GET is used.

Table 2 — Structure of TAP request using HTTP GET

| URL component | Description |
|---|---|
| http://host:port]/path[?[name[=value] {&name=[value]}]] | Base-URL (prefix) of service operation. [] denotes 0 or 1 occurrence of an optional part; {} denotes 0 or more occurrences. |
| name=value& | One or more standard request parameter name/value pairs as defined for each operation by this document. |

## 8.1.4 HTTP POST

TAP uses the "POST" method of the HTTP protocol (IETF RFC 2616 [11]) for submitting query parameters to both the /async and /sync endpoints and for modifying the state of a job on the /async endpoint. In addition, POST musty be used to upload a table inline as described in  2.5.2 . Parameters should be URL encoded in a POST whenever they would need to be URL encoded for a GET.

## 8.2 General HTTP response rules

Upon receiving a valid request, the server sends a response corresponding exactly to the request as detailed in section  2.7  of this document, or send a service exception if unable to respond correctly. Only in the case of Version Negotiation (see 2.8.2) may the server offer a differing result. Upon receiving an invalid request, the server returns an error document as described in section  2.7.3 .

A server may send an HTTP Redirect message (using HTTP response codes as defined in IETF RFC 2616 [11]) to an absolute URL that is different from the valid request URL that was sent by the client.  HTTP Redirect causes the client to issue a new HTTP request for the new URL.  Several redirects could in theory occur. Practically speaking, the redirect sequence ends when the server responds with a valid TAP response. The final response is a TAP response that corresponds exactly to the original request (or a service exception).

Response objects are accompanied by the appropriate Multipurpose Internet Mail Extensions (MIME) type (IETF RFC 2045 [12]) for that object by setting the *Content-Type* header in the HTTP response.  A list of MIME types in common use on the internet is maintained by the Internet Assigned Numbers Authority (IANA) . Allowable types for operation responses and service exceptions are discussed in  2.7.1 .

Response objects should be accompanied by other HTTP entity headers as appropriate and to the extent possible. In particular, the *Expires* and *Last-Modified* headers provide important information for caching; *Content-Length* may be used by clients to know when data transmission is complete and to efficiently allocate space for results, and *Content-Encoding* or *Content-Transfer-Encoding* may be necessary for proper interpretation of the results.